\documentstyle[prb,eqsecnum,aps,epsf,ifthen,multicol]{revtex}
\tighten
\newcommand{\wide}[2]{
\end{multicols}
\widetext \noindent \ifthenelse{\equal{#1}{t}} {} {
\raisebox{0.1in}[0in][0.02in]{$\rule{3.575in}{0.002in}
\rule{0.002in}{0.08in}$} } #2 \ifthenelse{\equal{#1}{b}} {} {
{\raisebox{-0.1in}[0in][0.02in]
{\hspace{3.575in}$\rule{0.002in}{0.08in}
\rule[0.08in]{3.575in}{0.002in}$} } }
\begin{multicols}{2}
\noindent }

\newcommand{\alphaq}{\alpha_{{\bf q}}}
\newcommand{\signum}{\mbox{sign}}

\begin{document}

\draft

\title{Superconducting fluctuations in granular metals with a large coupling between the grains}

\author{B.~S.~Skrzynski$^{(1)}$, I.~S.~Beloborodov$^{(2,3)}$ and K.~B.~Efetov$^{(1,4)}$}
\address{$^{(1)}$Theoretische Physik III, Ruhr-Universit\"{a}t
Bochum, 44780 Bochum, Germany, \\ $^{(2)}$ Bell Laboratories,
Lucent Technologies, Murray Hill, NJ 07974, \\ $^{(3)}$ Department
of Physics, University of Colorado, CB 390, Boulder, CO 80390, \\
$^{(4)}$ Landau Institute for Theoretical Physics, 117940 Moscow,
Russia}

\date{\today} \maketitle

\begin{abstract}
We study the fluctuation conductivity of superconducting granular
metals at low temperatures and strong magnetic field destroying
the Cooper pairs. Explicit calculations are performed for larger
values of the coupling between the grains than those considered in
previous works. We show that in a broad region of the coupling
constants the superconducting fluctuations still significantly
reduce the conductivity leading to a negative magnetoresistance.
\end{abstract}


\begin{multicols}{2}

\section{Introduction}
During the last decade study of electric properties of
non-homogeneous metals attracted a lot of attention. In
particular, granular metals have been investigated recently in a
number of experimental works \cite{kn:gerber,kn:gantmakher}.
Theory of superconducting fluctuations in the granulated
superconductors was suggested recently in
Ref.~\cite{kn:beloborodov,kn:beloborodov2}.

In these works a three-dimensional (3D) array of grains placed in
a strong magnetic field $H>H_c$, where $H_c$ is the field
destroying the superconducting gap in the grains, and at low
temperatures $T \ll T_c$, where $T_c$ is the superconducting
transition temperature, was considered. It was demonstrated that
the correction to the fluctuation conductivity is \emph{negative}
and this effect should exist even at zero temperature. The
resistivity increases and only at extremely strong magnetic
fields, $H \gg H_c$, reaches its classical value. Therefore the
system exhibits a \emph{negative magnetoresistance}. This theory
may explain existing experiments \cite{kn:gerber}.

It was important for the calculations presented in
Ref.~\cite{kn:beloborodov} that the dimensionless conductance
satisfied the condition $g \ll \Delta_0 / \delta$, here $\Delta_0$
is the BCS gap at zero magnetic field and $\delta$ is the mean
level spacing.

The same effect of the negative magnetoresistance was obtained in
a recent paper \cite{kn:galitski} for two-dimensional (2D)
homogeneous superconducting samples at low temperatures, $T \ll
T_c$, and strong magnetic field, $H>H_c$. The limit of the
homogeneous metal is opposite to the one considered in the works
\cite{kn:beloborodov,kn:beloborodov2} because the dimensionless
conductance $g$ of a homogeneous sample is proportional to $k_Fl$,
where $k_F$ and $l$ are the Fermi momentum and the mean free path,
respectively, and can be very large. The negative
magnetoresistance can also be seen under certain circumstances in
High-$T_c$ superconductors \cite{kn:ioffe,varlamov2,kn:varlamov1}
for high temperatures, $T>T_c$, and low magnetic fields, $H<H_c$.

In the present paper we generalize the results of
Ref.~\cite{kn:beloborodov,kn:beloborodov2} to larger values of the
tunneling dimensionless conductance $g$. In particular, the
assumption that the dimensionless conductance, $g$, is restricted
from above by $\Delta_0 / \delta$ is now dropped. This means that
the structure of the granular metal becomes more similar to that
of a bulk metal. The main question we are dealing with in this
paper is if the superconducting fluctuations may cause a negative
magnetoresistance in the granulated systems with the larger
coupling between the grains. At the same time, our region of
parameters is different from the limit of a homogeneous metal. We
assume that
\begin{equation} \label{eq:restriction}
1 \ll g \ll E_T / \delta,
\end{equation}
where $E_T= D/R^2$ is the Thouless energy of a single grain, $D$
is the diffusion coefficient and $R$ is the radius of the grain.
The last inequality in Eq.~(\ref{eq:restriction}), although being
more general than that used previously
\cite{kn:beloborodov,kn:beloborodov2}, means that the granular
structure is still important for our consideration.

The granular material we consider now consists of a 3D array of
metallic grains with a typical diameter of the grains of $100 \pm
20$\AA. The electrons can tunnel from one grain to another. It is
this tunneling that determines the properties of the entire
system. Inside the grains there can be impurities and the shape of
each grain is not perfect, so that the electrons are scattered
randomly by the boundaries. Since the hopping amplitude is not
very large, the macroscopic charge transfer is determined by the
ratio of the hopping amplitude $t$ to the mean level spacing
$\delta$, or, in other words, by the dimensionless conductance $g=
\case{\pi^2}{4} \left( \case{t}{\delta} \right)^2$. In the limit
$t \gg \delta$ the discreteness of the energy spectrum in a single
grain is not resolved and therefore the electron motion is
diffusive through many grains. This limit corresponds to a
macroscopically weak disorder and results in a large dimensionless
conductance $g \gg 1$. Below, we restrict our consideration by
this limit.

Let us discuss what happens with a granular metal at low
temperatures. Below the critical temperature, $T_c$, the
electron-phonon interaction leads to the formation of a
superconducting gap in each grain and Cooper pairs appear.
Applying a strong magnetic field one destroys the superconducting
gap in each grain and comes to the picture of a normal metal with
superconducting fluctuations. Our calculations are performed in
this regime.

We assume that the energy parameters are ordered as follows
\begin{equation} \label{eq:energyscales}
    \delta \ll t, \Delta_0 \ll E_T,
\end{equation}
The last inequality in Eq.~(\ref{eq:energyscales}) means that the
size of a single grain, $R$, is much smaller than the coherence
length $\xi_0$. In this limit the superconducting fluctuations in
a single grain are zero dimensional. We want to emphasize that all
energies are smaller than the Thouless energy, $E_T$, and as a
consequence the behavior of the system does not depend on grain
boundaries or on individual scattering processes. Due to the large
values of the conductance $g \gg 1$ we may neglect weak
localization and charging effects. Therefore all the effects
considered below are entirely due to the superconducting
fluctuations.

The superconducting pairing inside the grains can be destroyed by
both the orbital mechanism and the Zeeman splitting. The critical
magnetic field $H_{c}^{or}$ destroying the superconductivity in a
single grain in this case can be estimated as $H_{c}^{or}R\xi
\approx \phi _{0}$, where $\phi _{0}=hc/e $ is a flux quantum, $R$
is the radius of a single grain and $\xi =\sqrt{\xi _{0}l}$ is the
superconducting coherence length. The Zeeman critical magnetic
field $H_{c}^{z}$ can be written as $g\mu _{B}H_{c}^{z}=\Delta
_{0}$, where $\mu_B$ is Bohr's magneton and $g$ is the Land\'{e}
factor. The ratio of this two fields can be written in the form
$H_{c}^{or}/H_{c}^{z}\approx R_{c}/R$, where $R_{c}=\xi
(p_{0}l)^{-1}$. For $R>R_{c}$ the orbital critical magnetic field
is smaller than the Zeeman critical magnetic field
$H_{c}^{or}<H_{c}^{z}$ and the superconductivity is suppressed by
the orbital motion of electrons. Although the Zeeman mechanism can
be easily included in the present consideration, we consider now
only the orbital mechanism of the destruction of the
superconductivity. This limit is opposite to the one considered in
Ref.~\cite{Aleiner97}, where the Zeeman splitting was assumed to
be the main mechanism of destruction of the Cooper pairs.
 A broader region of the conductance $g$ used in the present paper
makes the calculation somewhat more difficult than previously
because one has to consider additional diagrams and calculate them
using more complicated expressions for integrands. The remainder
of the paper is organized as follows. In
Sec.~\ref{section:themodel} we formulate the model. In
Sec.~\ref{section:fluctuationconductivity} we discuss the
fluctuation conductivity of granular metals. In
Sec.~\ref{section:wlcorrection} we discuss the Weak Localization
correction to conductivity of granular metals. Our results are
summarized in the conclusion.

\section{The Model} \label{section:themodel}
We assume that the grains are packed in a 3D lattice surrounded by
an isulator. The grains are coupled with each other and therefore
the electrons can hop from one grain to another. Inside the grains
as well on the surface there can be impurities and the electrons
can interact with phonons. The Hamiltonian of the system can be
written in the form
\begin{equation} \label{eq:hamiltonian}
    \hat{H} = \hat{H}_0 +\hat{H}_T,
\end{equation}
here $\hat{H}_0$ describes a single grain with electron-phonon
interaction in the presence of a strong magnetic field and is
given by
\begin{equation} \label{eq:interaction}
    \hat{H}_0 = \sum_{i,k} E_{i,k} a_{i,k}^{\dag} a_{i,k}
    - |\lambda| \sum_{i,k,k^{\prime}} a_{i,k}^{\dag} a_{i,-k}^{\dag}
    a_{i,-k^{\prime}} a_{i,k^{\prime}} +\hat{H}_{imp},
\end{equation}
where $i$ stands for the number of the grain, $k \equiv ({\bf k},
\uparrow), -k \equiv (-{\bf k}, \downarrow)$. The quantity
$\lambda$ is an interaction constant and $\hat{H}_{imp}$ describes
the elastic interaction of the electrons with impurities. The
interaction in Eq.~(\ref{eq:interaction}) contains diagonal matrix
elements only. This form of the interaction can be used provided
the superconducting gap $\Delta_0$ satisfies the last inequality
in Eq.~(\ref{eq:energyscales}). The second term in
Eq.~(\ref{eq:hamiltonian}) describes tunneling of electrons from
grain to grain and is given by
\begin{equation}
    \hat{H}_T = \sum_{i,j,p,q} t_{i,j,p,q} a_{ip}^{\dag} a_{jq}
    \exp(i \frac{e}{c} {\bf A} {\bf d}_{ij}) + c.c,
\end{equation}
where ${\bf A}$ is the vector potential, ${\bf d}_{ij}$ is a
vector connecting the center of grain $i$ with the center of a
neighboring grain $j$ ($|{\bf d}_{ij}|=2R$). The operator
$a_{ip}^{\dag}$ is the creation-operator of an electron in grain
$i$ in the state $p$ and $a_{ip}$ is the annihilation operator of
an electron in grain $i$ in the state $p$.

\section{Fluctuation Conductivity} \label{section:fluctuationconductivity}
In this section we consider the conductivity of granular metals in
detail. The dc conductivity $\sigma$ is related to the operator of
the electromagnetic response as \cite{kn:abrikosov}:
\begin{equation} \label{eq:conductivity}
    \sigma=\lim_{\omega \rightarrow 0}
    \frac{Q^R(\omega)}{-i\omega},
\end{equation}
where $Q^R(\omega)$ is the analytical continuation of
$Q(i\omega_{\nu})$ into the upper complex half plane and is called
the \emph{retarded} operator of the electromagnetic response,
$\omega$ is the frequency of the external electromagnetic field.
In order to calculate $Q(i\omega_{\nu})$ we use Matsubara's
diagram technique \cite{kn:abrikosov}. After calculation of
$Q(i\omega_{\nu})$ for imaginary frequencies we have to carry out
the analytical continuation of $Q(i\omega_{\nu})$ into the region
of real frequencies: $i\omega_{\nu} \rightarrow \omega +i0^+$. All
diagrams which contribute to the conductivity of the granular
metal are shown in Fig.~\ref{fig:alldiagrams}. The same class of
diagrams describe the conductivity of the bulk metal.
\cite{kn:galitski,kn:altshuler3} Scattering of the electrons
inside the grains by impurities is included in the Born
approximation, giving rise to a scattering mean free time $\tau$
and resulting in a renormalization of the single electron normal
state Green's function to $ G^0 (i\varepsilon_n, {\bf
p})=(i\varepsilon_n -\xi({\bf p})+i/ 2\tau
\signum(\varepsilon_n))^{-1}$, here $\varepsilon_n= (2n+1)\pi T$
is the fermion frequency and $\xi({\bf p})= \varepsilon({\bf
p})-\varepsilon_F$ is the electron energy counted from the Fermi
level. For $l \ll L_c$, where $l$ is the mean free path and $L_c$
is the cyclotron radius, we can treat the Green's function in the
quasiclassical approximation. In this approximation the magnetic
field results in the appearance of an additional phase:
\begin{equation}
    G(i\varepsilon_n, {\bf r}- {\bf
    r}^{\prime})=G^0(i\varepsilon_n, {\bf r}-{\bf r}^{\prime})
    \exp \left(\frac{i e}{c} \int\limits_{{\bf r}}^{{\bf r}^{\prime}} {\bf A}
    d {\bf s} \right).
\end{equation}

\begin{figure}
    \epsfysize=1cm
    \centerline{\epsfbox{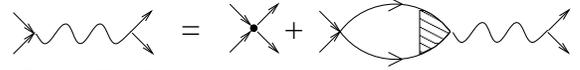}}
    \caption{\label{fig:propagator2} The Dyson equation in the ladder approximation for the
    propagator of the superconducting fluctuations. The black point represents the coupling constant
    and the shaded three-point vertex stands for the renormalized impurity vertex.}
\end{figure}

Each wavy line in the diagrams represents the propagator of the
superconducting fluctuations $K(i\Omega_k, {\bf q})$:
\begin{equation} \label{eq:propagator}
    K(i\Omega_k, {\bf q})=-\frac{1}{\nu_0} \left[ \ln \left(
    \frac{{\mathcal E}_0(H)+|\Omega_k|}{\Delta_0}
    \right)+\eta({\bf q}) \right]^{-1},
\end{equation}
here $\Omega_k =2k\pi T$ is the boson frequency, $\nu_0$ is the
density of states on the Fermi surface, $\eta ({\mathbf q})=
8/3\pi (g\delta/ \Delta_0) \sum_{i=1}^{3} (1-\cos(q_i d))$
describes the tunneling of electrons from grain to grain;
${\mathcal E}_0 (H)=(2/5) (\phi/ \phi_0)^2 E_T$, where $\phi$ is
the magnetic flux through the grain. The propagator of
superconducting fluctuations, Eq.~(\ref{eq:propagator}), is
presented by the sum of all diagrams with two incoming and two
outgoing lines in Fig.~\ref{fig:propagator2}.
\wide{m}{
\begin{figure}
        \epsfysize=10cm
        \centerline{\epsfbox{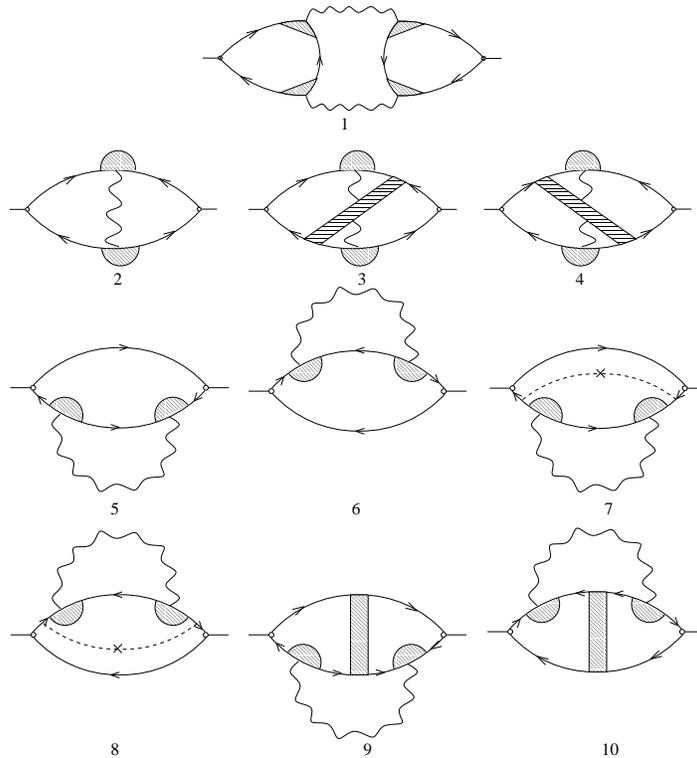}}
        \caption{\label{fig:alldiagrams} Diagrams for the leading order contribution
        to the fluctuation conductivity of granular metals. Wavy lines symbolize the propagator of the
        superconducting fluctuations, thin solid lines with arrows are the normal state Green's
        functions averaged over impurity positions and shaded semicircles are vertex corrections
        arising from impurities. Dashed lines with central crosses are additional impurity
        renormalizations and shaded blocks are impurity ladders. Diagram 1 is the
        \emph{Aslamazov-Larkin (AL)} contribution, diagram 2 is the \emph{Maki-Thompson (MT)},
        5, 6, 7 and 8 are the \emph{density of states (DOS)} diagrams.
        Diagrams 3,4 and 9,10 arise when one averages the DOS and MT diagrams over impurities.}
\end{figure}
}The impurity vertex entering these diagrams is presented as a
shaded half circle and has the form
\cite{kn:beloborodov,kn:beloborodov2}
\begin{eqnarray} \label{eq:impurityvertex}
    \lefteqn{\lambda (i\varepsilon_n, i\Omega_k-i\varepsilon_n, {\bf q}) =}  \nonumber \\
    & & \frac{1}{\tau} \frac{\theta(-\varepsilon_n(\Omega_k -
    \varepsilon_n))}{|2\varepsilon_n -\Omega_k| +{\mathcal E}_0
    (H) +\frac{16}{\pi} g \delta \sum_{i=1}^{3} (1-\cos(q_i d))},
\end{eqnarray}
where $\theta(x)$ is the Heaviside-function and the third term in
the denominator describes tunneling processes from grain to grain.
Each external vertex is given by $-2et{\bf d} \sin({\bf p d})$,
where $e$ is the charge of an electron, $t$ is the tunneling
amplitude and ${\bf d}$ is a vector connecting the centers of two
neighboring grains. In the following subsections we consider the
contributions to the fluctuation conductivity that arise from
these diagrams.

\subsection{Correction to the Conductivity due to Suppression of
DOS \label{sectiondos}} In this section we consider the correction
to conductivity due to suppression of the density of states (s.
diagrams 5 - 10). This (DOS) contribution arises from corrections
to the density of states due to the superconducting fluctuations.
Even at strong magnetic fields ($H>H_c$) there are still some
electrons that form fluctuational Cooper pairs. These electrons
are bound and cannot simultaneously take part in one-electron
charge transfer. This results in a reduction of the number of
carriers for the one-electron charge transfer and the conductivity
decreases. The analytical expression for the operator of the
electromagnetic response taking into account summation over the
spin indices has the form: \wide{m}{
\begin{equation} \label{eq:analyticaleqdos}
    Q(i\omega_{\nu})= \frac{8}{3} \sum_{i=1}^3 T \sum_{\Omega_{k}}
    \int \frac{d^3 {\bf q}}{(2 \pi)^3}
    K(i\Omega_{k}, {\bf q}) T\sum_{\varepsilon_n} C^2(i\varepsilon_n, i\Omega_k- i\varepsilon_n, {\bf q})
    I(i\Omega_{k}, i\varepsilon_{n+ \nu}),
\end{equation}
} where a factor of 2 originates from a similar diagram shown in
Fig.~\ref{fig:alldiagrams} which gives the same contribution to
conductivity. The analytical expression for $I(i\Omega_{k},
i\varepsilon_{n+ \nu})$ in Eq.~(\ref{eq:analyticaleqdos}) is given
by \wide{m}{
\begin{equation} \label{eq:block}
    I(i\Omega_{k}, i\varepsilon_{n+\nu})=
    \int \frac{d^3 {\bf q}^{\prime}}{(2\pi)^3}
    \sin^2(q_i^{\prime}d) \frac{4 e^2 t^2 d^2 V^2}{(2 \pi \nu_0\tau)^2}
    T \sum_{\varepsilon_n} \int \frac{d^3 {\bf p}_1 d^3 {\bf p}_2}{(2 \pi)^6}
    G(i\varepsilon_{n+\nu}, {\bf p}_1) G^2(i\varepsilon_n, {\bf p}_2)
    G(i\Omega_k -i\varepsilon_n, {\bf p}_2).
\end{equation}
}here ${\bf p}_1$ and ${\bf p}_2$ denote the momenta of electrons
inside the grains, ${\bf q}^{\prime}$ is the quasimomentum and
$\varepsilon_{n + \nu} = \varepsilon_n + \omega_{\nu}$. Inside the
Green's functions we may put $\Omega_k$ equal to zero because the
characteristic frequency of the superconducting fluctuation
propagator, $K(i\Omega_k, {\bf q})$, is of the order $\Omega_k
\sim \Delta_0$ which is much smaller than the Thouless energy
$E_T$. The integral over ${\bf p}_2$ in Eq.~(\ref{eq:block}) is
only nonzero when the poles of the Green's functions corresponding
to one grain lie on different sides of the axis of the real
numbers. Therefore $\varepsilon_n$ and $\Omega_k -\varepsilon_n $
must have different signs. In all other cases the result equals to
zero. For $\Omega_k, \varepsilon_n, \omega_{\nu} \ll 1/ \tau$ we
obtain for $I(i\Omega_k, i\varepsilon_{n+ \nu})$
\begin{equation}\label{eq:functioni}
    I(i\Omega_k, i\varepsilon_{n+ \nu})=\frac{4 g e^2}{\nu_0^2 \pi^2 d}
    \theta(-\varepsilon_n \varepsilon_{n+\nu})
    \theta(-\varepsilon_n(\Omega_k-\varepsilon_n)).
\end{equation}
Now using Eq.~(\ref{eq:functioni}) we calculate the sum over
$\varepsilon_n$ in Eq.~(\ref{eq:block})\wide{m}{
\begin{equation} \label{eq:functiond1}
    D(i\Omega_k, i\omega_{\nu}, {\bf q}) = T \sum_{\varepsilon_n}
    C^2(i\Omega_k, i\Omega_k -i\varepsilon_n , {\bf q})
    I(i\Omega_k, i\varepsilon_{n+ \nu})
    = \frac{16 g e^2}{d} T \sum_{n=-\nu}^{-1}
    \frac{\theta(-\varepsilon_n (\Omega_k-\varepsilon_n))}
    {(| 2\varepsilon_n -\Omega_k |+ 4\pi T \alphaq)^2},
\end{equation}
Carrying out the summation over the $\varepsilon_n$ in
Eq.~(\ref{eq:functiond1}) we obtain for the function $D(i\Omega_k,
i\omega_{\nu}, {\bf q})$ the following result
\begin{equation} \label{eq:functiond}
    D(i\Omega_k, i\omega_{\nu}, {\bf q})= -\frac{g e^2 }{\pi^2 d T}
    \theta(\Omega_k +\omega_{\nu})
    \left[\psi^{\prime} \left(\frac{1}{2}+\frac{2\omega_{\nu}+\Omega_k}{4\pi T}+\alphaq\right)
    -\psi^{\prime} \left(\frac{1}{2}+\frac{|\Omega_k|}{4\pi T}+\alphaq \right)
    \right],
\end{equation}
}where $\psi(x)$ is the logarithmic derivative of the
Gamma-function and $\alphaq$ is given by
\begin{equation} \label{eq:alphaq}
    \alphaq = \frac{1}{4\pi T} \left({\mathcal E}_0 (H) +\frac{16}{\pi}
    g \delta \sum_{i=1}^{3}(1-\cos(q_i d)) \right).
\end{equation}
The second term in Eq.~(\ref{eq:alphaq}) arises due to the
renormalization of the Cooperons when tunneling processes from
grain to grain are taken into account. We insert
Eq.~(\ref{eq:functiond}) into Eq.~(\ref{eq:analyticaleqdos}) and
present the operator of the electromagnetic response in the
following form:
\begin{equation} \label{eq:response}
    Q(i\omega_{\nu})=\frac{8}{3} \sum_{i=1}^3 T \sum_{\Omega_{k}} \int
    \frac{d^3 {\bf q}}{(2 \pi)^3} D(i\Omega_k, i\omega_{\nu},
    {\bf q}) K(i\Omega_k, {\bf q}).
\end{equation}
Now the function $Q(i\omega_{\nu})$ must be continued analytically
into the upper complex half-plane of the frequency. The analytical
continuation in Eq.~(\ref{eq:response}) is carried out in Appendix
\ref{sec:appdos}. As a result, we obtain for the operator of the
electromagnetic response $Q(i\omega_{\nu})$  after analytical
continuation the following expression \wide{m}{
\begin{equation} \label{eq:response2}
    Q^R(\omega) = \frac{- i\omega 2 g e^2}{\pi^4 d T^2 \Delta_0 \nu_0}
    \psi^{\prime \prime} \left( \case{1}{2}+ \case{\Delta_0}{4\pi T} \right)
     \sum_{i=1}^{3} \int \frac{d^3 {\bf q}}{(2\pi)^3}
     \left[ \int\limits_{0}^{\Omega_{max}} \coth \frac{\Omega}{2T}
     \frac{\Omega d\Omega}{ \frac{\Omega^2}{\Delta_0^2}+
    \tilde{\eta}^2({\bf q})} + \int\limits_{0}^{\infty} \frac{1}{\sinh^2 \frac{\Omega}{2T}}
    \frac{\Omega^2}{\frac{\Omega^2}{\Delta_0^2} + \tilde{\eta}^2({\bf q})} \frac{d \Omega}{2T} \right],
\end{equation}
where $\Omega_{max} \sim \Delta_0$ is an upper cut-off,
$\tilde{\eta}({\bf q})=\eta ({\bf q})+2h$ and
$h=\case{H-H_c}{H_c}$ is the reduced magnetic field. In
Eq.~(\ref{eq:response2}) we may put $\mathbf{q}=0$ inside
$\psi(x)$ because the main contribution to the integral over
$\mathbf{q}$ comes from small momentum and in this case $\alphaq$
is a slowly varying function of $\mathbf{q}$. The remaining
integrals over $\Omega$ in Eq.~(\ref{eq:response2}) can be easily
calculated and the final result for the electromagnetic response
is
\begin{equation} \label{eq:qdos2}
    Q^R(\omega)  = -i\omega \frac{2 g e^2 \Delta_0}{\pi^4 d T^2 \nu_0}
    \psi^{\prime \prime} \left(
    \frac{1}{2}+\frac{\Delta_0}{4\pi T} \right)
    \sum_{i=1}^{3} \int \frac{d^3 {\bf q}}{(2\pi)^3}
    \left[\ln \left(\frac{\xi}{\tilde{\eta}}\right) -  \frac{1}{2 \xi}
    - \psi (\xi) +
    \xi \psi^{\prime}(\xi)-1 \right],
\end{equation}
}here we introduced the dimensionless parameter $\xi= \Delta_0
\tilde{\eta} / 4\pi T$. For very low temperature $T \ll \Delta_0
\tilde{\eta}, (\xi \gg 1)$ we may use the asymptotic expansion for
$\psi(\xi)$ and finally obtain the correction to the conductivity
due to the suppression of the density of states:
\begin{equation} \label{eq:resultdos}
    \frac{\sigma^{DOS}}{\sigma_0} = -\frac{2}{\pi}
    \frac{\delta}{\Delta_0} \left \langle
    \ln \left( \frac{1}{\tilde{\eta}({\bf q})} \right) \right \rangle,
\end{equation}
where $\langle \ldots \rangle = V \int_{0}^{\frac{2\pi}{d}} \ldots
\frac{d^3 {\bf q}}{(2\pi)^3}$, and $\sigma_0 = \frac{8g e^2}{\pi
R}$ is the classical conductivity of a granular metal. One can see
that the DOS diagram gives a negative contribution to the
conductivity. The absolute value of $\sigma^{DOS}$ is a decreasing
function of the magnetic field $H$ and reaches its maximum value
at the critical magnetic field $H=H_c$. The absolute value of this
maximum can be estimated as \cite{kn:beloborodov,kn:beloborodov2}
\begin{equation}
    \left |\frac{\sigma^{DOS}_{max}}{\sigma_0} \right|
    \sim \frac{\delta}{\Delta_0} \ln \left(\frac{\Delta_0}{g\delta} \right).
\end{equation}
This maximum value is smaller than unity and this fact ensures our
diagrammatic expansion. The conductivity $\sigma^{DOS}$ is
independent of the temperature and therefore it remains finite in
the limit $T \rightarrow 0$. This fact indicates that there are
still virtual Cooper pairs even at zero temperature and strong
magnetic field. The quantity $\sigma^{DOS}$ becomes comparable
with $\sigma_0$ when $g$ is of the order of unity. Such values of
$g$ mean that we would not be far from the metal-insulator
transition. In this case we have to take into account all
localization effects and Eq.~(\ref{eq:resultdos}) can be used only
for rough estimates.

We would like to note that Eq.~(\ref{eq:resultdos}) does not
differ from those written in
Refs.~\cite{kn:beloborodov,kn:beloborodov2}. However, other
contributions to the conductivity considered in the next
subsections may change, so the extension of the calculations to
the entire region specified by Eq. (\ref{eq:restriction}) is not
as simple.
\begin{figure}
    \epsfysize=4cm
    \centerline{\epsfbox{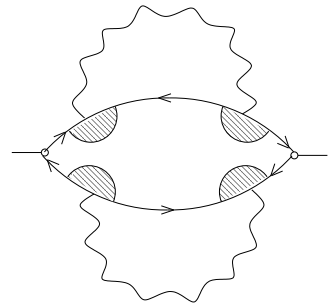}}
    \caption{\label{fig:higherorderdos} Higher order correction to the DOS}
\end{figure}
As it was shown in Ref.~\cite{kn:beloborodov,kn:beloborodov2} we
can neglect the higher order corrections to the DOS, an example is
shown in Fig.~\ref{fig:higherorderdos}. At low temperatures $T \ll
\Delta_0 \tilde{\eta}$ this diagram contains an additional small
factor of $(\delta / \Delta_0) \ln \tilde{\eta}$.

In the following sections we discuss the Aslamazov-Larkin (AL) and
Maki-Thompson (MT) contributions to the fluctuation conductivity.
It turns out that the AL contribution is proportional to $T^2$
whereas the MT contribution can be divided into two parts. One is
proportional to $T^2$ at low temperatures and the other one
remains finite when $T \rightarrow 0$.

\subsection{Aslamazov-Larkin Contribution to Conductivity}
This contribution originates from the ability of virtual Cooper
pairs to carry an electrical current \cite{kn:aslamazov2}. The
diagram for the operator of the electromagnetic response is
represented by the first diagram in Fig.~\ref{fig:alldiagrams} and
its analytical expression has the following form \wide{m}{
\begin{equation} \label{eq:al1}
    Q^{AL}(i\omega_{\nu})= \frac{4}{3} \sum_{i=1}^{3} T
    \sum_{\Omega_k} \int \frac{d^3 {\bf q}}{(2\pi)^3 } K(i\Omega_k,
    {\bf q}) K(i\Omega_k -i\omega_{\nu}, {\bf q}) B^2(i\Omega_k,
    i\omega_{\nu}, {\bf q}),
\end{equation}
$B(i\Omega_k , i\omega_{\nu}, {\bf q})$ describes the block of the
Green's functions:
\begin{eqnarray} \label{eq:functionb}
    \lefteqn{B(i\Omega_k,i\omega_{\nu},{\bf q}) = -i \int \frac{d^3
    {\bf q}^{\prime}}{(2 \pi )^3} \sin(q_i^{\prime}d) \cos((q_i-q_i^{\prime})d)
    \frac{e t^2 d V^2}{(\pi \nu_0 \tau)^2}
    T \sum_{\varepsilon_n} \int \frac{d^3 {\bf p}_1 d^3
    {\bf p}_2}{(2\pi)^6}} \nonumber \\
    && \times G(i\varepsilon_{n+\nu}, {\bf p}_1)
    G(i\Omega_k -i\varepsilon_{n+\nu}, {\bf p}_1)
    G(i\varepsilon_n, {\bf p}_2) G(i\Omega_k- i\varepsilon_{n+\nu}, {\bf p}_2)
    C(i\varepsilon_{n+\nu}, i\Omega_k-i\varepsilon_{n+\nu},
    {\bf q}) C(i\varepsilon_n, i\Omega_k-i\omega_{\nu}, {\bf q}),
\end{eqnarray}
}here $\sin(q_i^{\prime} d)$ denotes the current-vertex and
$\cos((q_i -q_i^{\prime})d)$ describes the tunneling-vertex. In
this integral we may put $\Omega_k=0$ and $\omega_{\nu}=0$ because
in the vicinity of the critical magnetic field $H_c$ the leading
contribution to the response $Q^{AL}$ arises from the fluctuation
propagators rather than from the frequency dependence of the
function $B$, Eq.~(\ref{eq:functionb}). Therefore we neglect the
dependence of function $B$ on frequencies $\Omega_k$ and
$\omega_{\nu}$. Integrating over the momenta ${\bf p}_1, {\bf
p}_2$ and quasimomentum ${\bf q}^{\prime}$, then summing over the
internal frequency $\varepsilon_n$ we obtain in the limit of low
temperature for function $B({\bf q})$ the result:
\begin{equation} \label{eq:loopal}
    B({\bf q})=-i \frac{32 g e d}{\pi \Delta_0}
    \int \sin(q_i^{\prime} d) \cos((q_i- q_i^{\prime})d) \frac{d^3 {\bf q}^{\prime}}{(2\pi)^3}.
\end{equation}
As in Sec.~\ref{sectiondos}, in order to calculate the response
$Q^{AL}$, we have to make the analytical continuation from the
Matsubara frequency $\omega_{\nu}$ to real values of $\omega$.
This can be achieved by transforming the sum over $\Omega_k$ in
Eq.~(\ref{eq:al1}) into a contour integral. This procedure allows
to carry out the analytical continuation of $Q(i\omega_{\nu})$
into the upper complex half plane, $i\omega_{\nu} \rightarrow
\omega +i0^+$. As a result for the sum over the $\Omega_k$ in
Eq.~(\ref{eq:al1}) we obtain
\begin{eqnarray}\label{eq:sumoveromegak3}
    T \sum_{\Omega_k} K(i\Omega_k, {\bf q}) K(i\Omega_k -
    i\omega_{\nu}, {\bf q}) \rightarrow -i\omega \frac{2 \pi T^2}{3\nu_0^2 \Delta_0^2}
    \frac{1}{\tilde{\eta}^4}.
\end{eqnarray}
Using Eq.~(\ref{eq:loopal}) and Eq.~(\ref{eq:sumoveromegak3}) we
finally obtain for the AL contribution to the fluctuation
conductivity
\begin{equation}\label{eq:resultal}
    \frac{\sigma^{AL}}{\sigma_0}= \frac{64}{27} g \frac{\delta^2 T^2}{\Delta_0^4}
    \sum_{i=1}^{3} \left \langle
    \frac{\sin^2 (q_i d)}{\tilde{\eta}^4({\bf q})} \right \rangle,
\end{equation}
which agrees with the result obtained in
Refs.~\cite{kn:beloborodov1,kn:beloborodov2}. One can see that
$\sigma^{AL}$ gives a positive contribution to the fluctuation
conductivity and is proportional to $T^2$, therefore it vanishes
in the limit $T \rightarrow 0$.

Let us estimate the right hand side in Eq.~(\ref{eq:resultal}) for
the case of low temperatures, $T \ll \Delta_0 \tilde{\eta}$, and
near the critical magnetic field, $h \ll g\delta/ \Delta_0$. The
main contribution to the integral in Eq.~(\ref{eq:resultal}) comes
from small momenta. Therefore we can make an expansion in ${\bf q
}$ of $\sin (q_i d)$ and $\cos(q_i d)$. Retaining only the first
nonvanishing terms and extending the range of integration from
$2\pi /d$ to infinity we obtain
\begin{eqnarray}
    \frac{\sigma^{AL}}{\sigma_0}
    \sim g^{-3/2} \frac{ T^2}{ \Delta_0^{3/2}\delta^{1/2}} \left(\frac{H_c}{H-H_c}\right)^{3/2}.
\end{eqnarray}
The AL contribution grows when approaching the critical magnetic
field $H_c$ thus leading to a decrease of resistivity. In order to
determine which contribution to the conductivity will dominate we
compare $\sigma^{AL}$ with the maximum value of $\sigma^{DOS}$
\begin{equation}\label{eq:comparison}
    \left | \frac{\sigma^{AL}}{\sigma^{DOS}_{max}}\right|
    \sim  \frac{g^{-3/2} T^2}{ \Delta_0^{1/2}\delta^{3/2}}
    \ln^{-1} \left( \frac{\Delta_0}{g \delta} \right)
    \left( \frac{H_c}{H-H_c}\right)^{3/2}.
\end{equation}
For $h\ll g\delta / \Delta_0$ and sufficiently low temperatures,
$T \ll \Delta_0 \tilde{\eta}$, one can see from
Eq.~(\ref{eq:comparison}) that $|\sigma^{AL}/ \sigma^{DOS}| \ll
1$. This means that the AL contribution cannot change the
monotonous increase of the resistivity of granular metals when
decreasing the magnetic field.

\subsection{Maki-Thompson Contribution to Conductivity}
This contribution to the conductivity comes from coherent electron
scattering forming a Cooper pair on impurities \cite{kn:maki},
\cite{kn:thompson}. The MT contribution is represented by the
second diagram in Fig.~\ref{fig:alldiagrams}. The analytical
expression for the operator of the electromagnetic response for
the MT contribution to fluctuation conductivity is given by:
\begin{equation} \label{eq:analyticalexpressionmt}
    Q^{MT}(i\omega_{\nu})= \frac{2}{3} \sum_{i=1}^{3}
    T \sum_{\Omega_k} \int \frac{d^3{\bf q}}{(2\pi)^3}
    K(i\Omega_k, {\bf q}) B(i\Omega_k, i\omega_{\nu}, {\bf q}),
\end{equation}
where $K(i\Omega_k, {\bf q})$ is the propagator of the
superconducting fluctuations and $B(i\Omega_k, i\omega_{\nu}, {\bf
q})$ is a function describing the contribution of the loop. This
function has the form \wide{m}{
\begin{eqnarray} \label{eq:loop}
    \lefteqn{B(i\Omega_k, i\omega_{\nu}, {\bf q}) =
    \int \sin(q_{i}^{\prime} d) \sin((q_i -q_{i}^{\prime})d)
    \frac{d^3 {\bf q}^{\prime}}{(2 \pi)^3} \frac{e^2 t^2 d^2 V^2}{(\pi \nu_0 \tau)^2}
    T\sum_{\varepsilon_n} \int \frac{d^3 {\bf p}_1 d^3
    {\bf p}_2}{(2 \pi)^6}} \nonumber \\
    && \times G(i\varepsilon_n, {\bf p}_1) G(i\Omega_k -i\varepsilon_n , {\bf p}_1)
    G(i\varepsilon_{n -\nu}, {\bf p}_2)
    G(i\Omega_k - i\varepsilon_{n -\nu}, {\bf p}_2)
    C(i\varepsilon_n, i\Omega_k -i\varepsilon_n,
    {\bf q}) C(i\varepsilon_{n -\nu},i\Omega_k- i\varepsilon_{n -\nu}, {\bf q}).
\end{eqnarray}
}In evaluating the sum over the Matsubara frequency
$\varepsilon_n$ it is useful to break up the sum into two parts.
In the first part $\varepsilon_n$ is in the domains $]-\infty,
-\omega_{\nu}[$ and $[0, \infty[$. This gives rise to the
\emph{regular part} of the MT diagram. The second
(\emph{anomalous}) part of the MT diagram arises from the
summation over the $\varepsilon_n$ in the domain
$[-\omega_{\nu},0[$. Using this we can perform the sum over the
$\varepsilon_n$ and after integration over the momenta we can
write the function $B$ as a sum of an \emph{anomalous} $B^{an}$
and a \emph{regular} $B^{reg}$ contribution to the MT diagram:
\begin{equation}
    B(i\Omega_k, i\omega_{\nu}, {\bf q})=-32 \frac{g e^2}{d} \cos(q_i d)
    (B^{an}+B^{reg}),
\end{equation}
where \wide{m}{
\begin{equation}
    \label{eq:banomalous}
    B^{an}  =  -\frac{1}{4\pi} \frac{\theta(\Omega_k) \theta(\omega_{\nu}-\Omega_{k+1})}{\omega_{\nu}+ 4\pi T \alphaq}
    \left[\psi \left(\frac{1}{2}+\frac{2\omega_{\nu}-\Omega_k}{4\pi T}+\alphaq\right)
    -\psi\left( \frac{1}{2}+\frac{\Omega_k}{4\pi T}+\alphaq\right)\right],
\end{equation}
\begin{equation} \label{eq:bregular}
    B^{reg} = \frac{1}{8\pi \omega_{\nu}} \left[\psi \left(\frac{1}{2}+
    \frac{2\omega_{\nu}+\Omega_k}{4\pi T}+\alphaq\right)-
    \psi \left( \frac{1}{2}+\frac{\Omega_k}{4\pi T}+\alphaq \right) \right].
\end{equation}
}Note that the anomalous and regular part have different signs and
the anomalous part has an additional diffusion pole in comparison
with the regular one.

\subsubsection{The Anomalous MT Contribution to Conductivity}
The term $B^{an}$ corresponds to the \emph{anomalous} MT
contribution, it appears in the case of a special pole arrangement
in the integration over momenta in Eq.~(\ref{eq:loop}), when the
poles of the Green's functions corresponding to different grains
lie on different sides of the axis of the real numbers. In order
to calculate the fluctuation conductivity we have to insert
$B^{an}$ into Eq.~(\ref{eq:analyticalexpressionmt}). Then, one
should perform the analytical continuation of
$Q^{an}(i\omega_{\nu})$ to real values of the external frequency
$\omega$. Finally, using Eq.~(\ref{eq:conductivity}) we get the
conductivity. To this end we represent $Q^{an}$ in the following
form:
\begin{equation} \label{eq:qanomalous}
    Q^{an}(i\omega_{\nu})=- \frac{16 g e^2}{3 \pi d \nu_0}
    \sum_{i=1}^3 \int\frac{d^3 {\bf q}}{(2\pi)^3}
    \frac{\cos(q_id) F(\omega_{\nu}, {\bf q})}{\omega_{\nu}+4\pi T \alphaq}.
\end{equation}
The function $F(i\omega_{\nu}, {\bf q})$ is given by
\cite{kn:aslamazov}:
\begin{equation}\label{eq:functionf}
    F(i\omega_{\nu}, {\bf q})=T
    {\sum_{\Omega_k=0}^{\omega_{\nu-1}}}^\prime f(\Omega_k,
    \omega_{\nu}, {\bf q}),
\end{equation}
where $f(i\Omega_k, i\omega_{\nu}, {\bf q})=
    \case{\psi \left(\frac{1}{2}+\frac{2\omega_{\nu}-\Omega_k}{4\pi T}
    +\alphaq \right)- \psi \left( \frac{1}{2}+\frac{\Omega_k}{4\pi T}
    +\alphaq \right)}{2h+\frac{|\Omega_k|}{\Delta_0}+\eta({\bf q})}$.
Because of the presence of the Heaviside function $\theta$ in
Eq.~(\ref{eq:banomalous}), $B^{an}$ is only nonzero in the domain
$[\Omega_1, \omega_{\nu -1}]$. The upper limit of the sum over
$\Omega_k$ in Eq.~(\ref{eq:functionf}) depends on the external
frequency $\omega_{\nu}$. Therefore it is not correct simply to
make the substitution $i\omega_{\nu} \rightarrow \omega$. Note
that we have calculated Eq.~(\ref{eq:banomalous}) only for
$\Omega_k \geq 0$ but one can easily obtain the corresponding
expression for $\Omega_k <0$ replacing $\Omega_k$ by $|\Omega_k|$.
Therefore, instead of summing over all values of $\Omega_k$ we
extract the term $\Omega_k=0$ and multiply the sum over $\Omega_k
>0$ by a factor of 2. The prime on the sum in Eq.~(\ref{eq:functionf}) indicates that the term with
$\Omega_k =0$ should be multiplied by 1/2. The analytical
continuation is achieved by transformation of the sum into a
contour integral. The final result has the following form
\cite{kn:aslamazov}: \wide{m}{
\begin{equation} \label{eq:mtall}
    F^R(\omega, {\bf q}) = -\frac{i\omega}{4\pi T} \int_{0}^{\infty} \frac{d
    \Omega}{\sinh^2 \frac{\Omega}{2T}}
    \frac{\psi \left( \frac{1}{2}+\frac{i\Omega}{4\pi T} +\alphaq \right)
    - \psi \left( \frac{1}{2} -\frac{i\Omega}{4\pi T} +\alphaq \right)}{2h -\frac{i \Omega}{\Delta_0}
    +\eta ({\bf q})}.
\end{equation}
}For $\Omega \ll 1$ we expand the numerator in a Taylor-series
around $\Omega=0$ and confine ourselves to the first nonvanishing
term. Then we obtain
\begin{equation}
    F^R(\omega, {\bf q})=\frac{2 i\omega}{(4\pi T)^2 \Delta_0}
    \psi^{\prime} \left( \case{1}{2} + \case{\Delta_0}{4\pi T} \right)
    \int\limits_{0}^{\infty}
    \frac{d\Omega}{\sinh^2 \case{\Omega}{2T}} \frac{\Omega^2}{\frac{\Omega^2}{\Delta_0^2} +\tilde{\eta}^2}.
\end{equation}
The last integral can be calculated in a similar way as the second
integral in Eq.~(\ref{eq:response2}) in Sec.~{\ref{sectiondos}}
and we obtain for the anomalous MT contribution to conductivity
the following result:
\begin{equation}\label{eq:resultmt}
    \frac{\sigma_{an}^{MT}}{\sigma_0}=
    \frac{4 \pi \delta T^2}{9 \Delta_0^2} \sum_{i=1}^{3} \left \langle
    \frac{\cos(q_i d)}{4\pi T \alphaq \tilde{\eta}^2({\bf q})} \right \rangle.
\end{equation}

Let us estimate the anomalous MT contribution to the conductivity.
The main contribution to the integral in Eq.~(\ref{eq:resultmt})
comes from small momenta ${\bf q}$, therefore for $h \ll g\delta/
\Delta_0$ we expand the numerator and denominator in powers of
${\bf q}$. Confining ourselves to the first nonvanishing order in
${\bf q}$, we obtain the following result:
\begin{eqnarray} \label{eq:comparisonmt}
    \frac{\sigma_{an}^{MT}}{\sigma_0}
    &\sim & g^{-3/2} \frac{T^2}{\Delta_0^{3/2} \delta^{1/2}} \left( \frac{H_c}{H-H_c} \right)^{1/2}.
\end{eqnarray}
From Eq.~(\ref{eq:comparisonmt}) we see that $\sigma_{an}^{MT}$
gives a positive contribution to the fluctuation conductivity and
grows when approaching the critical magnetic field $H \rightarrow
H_c$. The anomalous MT contribution is proportional to $T^2$ as
temperature goes to zero. At zero temperature the anomalous MT
contribution vanishes.

It is interesting to compare $\sigma_{an}^{MT}$ with the
contribution to conductivity that arises from the suppression of
the density of states $\sigma^{DOS}$. The ratio of these two
quantities is given by
\begin{equation}
    \left| \frac{\sigma_{an}^{MT}}{\sigma_{max}^{DOS}} \right| \sim
     \frac{g^{-3/2} T^2}{\Delta_0^{1/2} \delta^{3/2}}
    \ln^{-1} \left(\frac{\Delta_0}{g \delta} \right)
    \left(\frac{H_c}{H-H_c} \right)^{1/2}.
\end{equation}
One can see that at $T \ll \Delta_0 \tilde{\eta}$ the anomalous MT
contribution is small compared to $\sigma_{max}^{DOS}$. Thus, we
conclude that the anomalous MT contribution cannot change the
monotonous increase of the resistivity of a granular
superconductor when decreasing the magnetic field.

\subsubsection{The Regular MT Contribution to Conductivity}
Let us now investigate the contribution to the conductivity
arising from the \emph{regular} part of the MT diagram. The
operator for the electromagnetic response can be written as
\wide{m}{
\begin{equation} \label{eq:qreg2}
    Q^{reg}(i\omega_{\nu})=
    \frac{8 g e^2}{3\pi d \omega_{\nu} \nu_0} \nonumber
    \sum_{i=1}^{3} T \sum_{\Omega_k} \int \frac{d^3 {\bf q}}{(2\pi)^3} \cos(q_i d)
     \frac{\psi \left(\frac{1}{2}+
    \frac{2\omega_{\nu}+\Omega_k}{4\pi T}+\alphaq\right)-
    \psi \left( \frac{1}{2}+\frac{\Omega_k}{4\pi T}+\alphaq \right)}
    {2h -\frac{i\Omega}{\Delta_0} +\eta({\bf q})}.
\end{equation}
}As before, we can write this sum as a contour integral. Then, we
have to perform the analytical continuation to real values of the
external frequency $i\omega_{\nu}$: $i\omega_{\nu} \rightarrow
\omega$. We expand the resulting expression in a Taylor-series up
to the second order in $\omega$. The static term is canceled by a
similar term in $Q(0)$. As a result we obtain for the operator of
the electromagnetic response:
\begin{equation}\label{eq:qreg3}
    Q_{reg}^{MT}= \frac{-2 i\omega ge^2   }{9 \pi^4 d T^2 \nu_0 \Delta_0}
    \psi^{\prime \prime} \left( \case{1}{2} +\case{\Delta_0}{4\pi T} \right)
    \int\limits_{0}^{\Omega_{max}} \coth \frac{\Omega}{2T}
    \frac{\Omega d \Omega}{\frac{\Omega^2}{\Delta_0^2} +\tilde{\eta}^2}.
\end{equation}
Performing the integration in Eq.~(\ref{eq:qreg3}) we obtain for
the regular MT contribution to conductivity in the low temperature
limit:
\begin{equation} \label{eq:sigmamtreg}
    \frac{\sigma_{reg}^{MT}}{\sigma_0}=-\frac{2}{9 \pi} \frac{\delta}{\Delta_0}
    \sum_{i=1}^{3} \left\langle \cos(q_i d) \ln \left( \frac{1}{\tilde{\eta}({\bf q})} \right) \right\rangle.
\end{equation}
Let us estimate the integral in the right hand side of
Eq.~(\ref{eq:sigmamtreg}). At very low temperatures, $T \ll
\Delta_0 \tilde{\eta}$, and near the critical magnetic field
$H_c$, $h \ll g\delta/ \Delta_0$, it turns out that
\begin{equation}
    \frac{\sigma_{reg}^{MT}}{\sigma_0} \sim -\frac{\delta}{\Delta_0}.
\end{equation}
The regular MT part gives a negative contribution to the
fluctuation conductivity and it is independent of temperature, so
that even at zero temperature $\sigma_{reg}^{MT}$ remains finite
and reduces the conductivity. We have shown before that at very
low temperature neither $\sigma^{AL}$ nor $\sigma_{an}^{MT}$ can
change the monotonous increase of resistivity caused by the
suppression of the density of states and the regular MT
contribution.

\subsection{Diagrams No. 3, 4, 7, 8 and 9,10}
\begin{figure}
    \epsfysize=4cm
        \centerline{\epsfbox{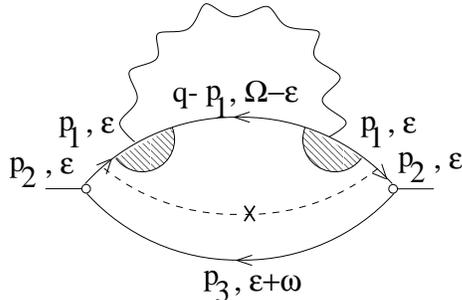}}
        \caption{\label{fig:mtdostype} DOS-type diagram with an additional impurity
        renormalization.}
\end{figure}

These diagrams arise when the DOS and MT diagrams are averaged
over the impurity positions. This averaging process results in the
appearance of an additional Cooperon connecting two different
grains with each other. Let us consider the diagram 9 in
Fig.~\ref{fig:alldiagrams}, the diagram 10 can be calculated in
the same way. The analytical expression of the operator of the
electromagnetic response reads as follows
\begin{equation} \label{eq:odresponse}
    Q(i\omega_{\nu})=\frac{4}{3} \sum_{i=1}^{3} T \sum_{\Omega_k}
    \int \frac{d^3 {\bf q}}{(2\pi)^3} K(i\Omega_k, {\bf q})
    B(i\Omega_k, i\omega_{\nu}, {\bf q}),
\end{equation}
where $K(i\Omega_k, {\bf q})$ is the propagator of the
superconducting fluctuations and $B(i\Omega_k, i\omega_{\nu}, {\bf
q})$ corresponds to the contribution of the loop. A factor of 2 in
Eq.~(\ref{eq:odresponse}) comes from the summation over spin
indices and another factor of 2 originates from a similar diagram
shown in Fig.~\ref{fig:alldiagrams}. The analytical expression for
the loop can be written as \wide{m}{
\begin{eqnarray}
    \lefteqn{B(i\Omega_k, i\omega_{\nu}, {\bf q}) = 4 \int
    \sin(q_i^{\prime} d)\frac{d^3 {\bf q}^{\prime}}{(2\pi)^3}
    \int \sin(q_i^{\prime \prime} d) \frac{d^3 {\bf q}^{\prime \prime}}{(2\pi)^3}
    \frac{e^2 t^2 d^2 V^2}{(2\pi \nu_0 \tau)^3}
    T \sum_{\varepsilon_n} \int \frac{d^3 {\bf p}_1 d^3 {\bf p}_2}{(2 \pi)^6}} \nonumber \\
    && \times G(i\varepsilon_{n +\nu}, {\bf p}_1) G^2(i\varepsilon_n, {\bf p}_2)
    G(i\Omega_k-i\varepsilon_n, {\bf p}_2)
    C^2(i\varepsilon_n, i\Omega_k- i\varepsilon_n, {\bf q})
    C(i\varepsilon_{n+ \nu}, i\Omega_k -i\varepsilon_n, {\bf q}),
\end{eqnarray}
}where ${\bf p}_1$ and ${\bf p}_2$ denote the momenta in the
different granules and ${\bf q}^{\prime}, {\bf q}^{\prime \prime}$
are quasimomenta. As before we can set $\Omega_k$ inside the
Green's functions equal to zero. We can immediately see that after
integration over ${\bf q}^{\prime}, {\bf q}^{\prime \prime}$ the
function $B(i\Omega_k, i\omega_{\nu}, {\bf q})$ is equal to zero
and the contribution to the fluctuation conductivity from this
diagram vanishes. The same reason holds for the diagram 3 and 4 in
Fig.~\ref{fig:alldiagrams} which come from the averaging of the
MT-diagram over impurity positions. Also the diagrams 7 and 8 of
Fig.~\ref{fig:alldiagrams} do not contribute to the fluctuation
conductivity. For simplicity we present the diagram 7 once again
in Fig.~\ref{fig:mtdostype}.
One can easily see that the corresponding analytical expression of
such a diagram contains a term of the following form $\int
\frac{d^3 {\bf p}_2}{(2\pi)^3} G^A(\varepsilon,{\bf p}_2)
G^A(\varepsilon,{\bf p}_2)=0$. Both Green functions have poles on
the same side of the complex plane and therefore the integral is
equal to zero. Thus, we may neglect diagrams 7 and 8. As we have
shown, only the diagrams 1 (AL), 2 (MT), 5 and 6 (DOS) make a
contribution to the fluctuation conductivity of granular metals.

\subsection{Final Formulae}
The calculations presented in the previous subsections show that
in granular superconducting metals fluctuations make a
considerable contribution to the conductivity in the normal phase
at low temperatures and strong magnetic field. The fluctuation
conductivity of granular metals is given by the DOS,
Aslamazov-Larkin and Maki-Thompson contribution
\begin{equation}
    \sigma^{fl}=\sigma^{DOS}+\sigma^{AL}+\sigma^{MT}.
\end{equation}
Other contributions from the diagrams in
Fig.~\ref{fig:alldiagrams} are equal to zero. At low temperatures,
$T \ll \Delta_0 \tilde{\eta}$, the final result for the total
fluctuation conductivity of granular metals can be written as:
\wide{m}{
\begin{eqnarray} \label{eq:fluctuationconductivity}
    \frac{\sigma^{fl}}{\sigma_0} = -\case{\delta}{\Delta_0}
    \left[
     \case{1}{2\pi} \left\langle  \ln \left( \case{1}{\tilde{\eta}({\bf q})} \right) \right\rangle
    -\sum_{i=1}^3 \left(
    \case{64}{27} g \case{\delta^2 T^2}{\Delta_0^4} \left \langle \case{\sin^2(q_i d)}{{\tilde{\eta}}^4 ({\bf q})}
    \right \rangle  +\case{4 \pi}{9} \case{T^2}{\Delta_0^2} \left \langle \case{\cos(q_i d)}{4\pi T \alphaq {\tilde{\eta}}^2({\bf q})}
    \right \rangle -\case{2}{9 \pi} \left \langle \cos(q_i d) \ln \left( \case{1}{\tilde{\eta}({\bf q})} \right) \right \rangle
    \right)
    \right].
\end{eqnarray}
}The Eq.~(\ref{eq:fluctuationconductivity}) is the main result of
our paper. For low temperatures, $T \ll \Delta_0 \tilde{\eta}$ and
strong magnetic field, $H>H_c$ the main contribution comes from
the first and from the last terms in the brackets in
Eq.~(\ref{eq:fluctuationconductivity}). As a consequence, the
superconducting fluctuation contribution to the conductivity of
granular metals is \emph{negative}. The quantitative result of
Eq.~ (\ref{eq:fluctuationconductivity}) is written for the region
$1 \ll g \ll \case{E_T}{\delta}$. In the next section we show that
the correction to the conductivity due to weak localization
effects can be neglected.

\section{Weak Localization Correction to Conductivity of Granular Metals} \label{section:wlcorrection}
\begin{figure}
        \epsfysize=3cm
        \centerline{\epsfbox{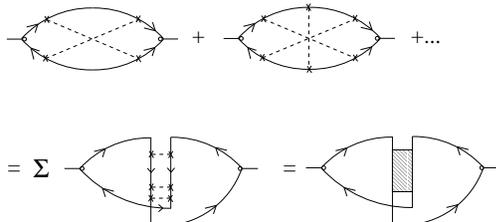}}
        \caption{\label{fig:weaklocalization}
        Weak localization correction to conductivity.
        The shaded block denotes the renormalized Cooperon.}
\end{figure}

The \emph{Weakly Localized Regime (WLR)} is the regime where
interference effects between different plane waves, being treated
independently, start to play a role. The interference of plane
waves leads to an increase of the probability to find an electron
at a certain place and this effect results in a reduction of the
conductivity. The diagram corresponding to the WL correction to
conductivity \cite{kn:gorkov} is shown in
Fig.~\ref{fig:weaklocalization}.
For the WL correction of granular metals
\cite{kn:beloborodov,kn:beloborodov2} we obtain
\begin{equation}\label{eq:wlcorrection}
    \delta \sigma^{WL}=-\frac{16}{3} g e^2 \delta d^2 \sum_{i=1}^{3} \int \frac{C(0, {\bf q})}{2\pi \nu_0}
    \cos(q_i d) \frac{d^3 {\bf q}}{(2\pi)^3},
\end{equation}
here $C(0, {\bf q})$ is the Cooperon taken at the frequency
$\omega=0$ and quasimomentum ${\bf q}$. Using
Eq.~(\ref{eq:impurityvertex}) and making the integration over
quasimomentum ${\mathbf q}$ in Eq.~(\ref{eq:wlcorrection}) we
obtain the final result for the WL correction to conductivity of
granular metals
\begin{equation}
    \frac{\delta \sigma^{WL}}{\sigma_0}
    \sim -g^{-3/2} \left(\frac{\Delta_0}{\delta}\right)^{1/2}.
\end{equation}
Similar to the case of a homogeneous sample, the WL correction to
conductivity has the negative sign.

Now let us compare the WL correction with $\sigma^{DOS}$. The
ratio of these two quantities is given by
\begin{equation}
   \left| \frac{\delta \sigma^{WL}}{\sigma^{DOS}_{max}} \right|
   = g^{-3/2} \left(\frac{\Delta_0}{\delta} \right)^{3/2}
   \ln^{-1} \left( \frac{\Delta_0}{g \delta} \right).
\end{equation}
For large $g \gg 1$, this ratio is small and therefore the WL
correction is always smaller than the correction which arises from
the suppression of the density of states.

\section{Conclusion}
We have obtained the fluctuation conductivity of granular metals
at low temperatures and strong magnetic field. Our main result is
given by Eq.~(\ref{eq:fluctuationconductivity}). To obtain this
result we assumed that the dimensionless conductance, $g$,
satisfies the inequality (\ref{eq:restriction}). Therefore, the
granular structure of the metal is essential for our
consideration. We have generalized the previous studies
\cite{kn:beloborodov,kn:beloborodov2} to the entire region $1 \ll
g \ll \case{E_T}{\delta}$. This case is still different from the
case of 2D homogeneous superconductor \cite{kn:galitski} where the
dimensionless conductance was assumed to be very large (of order
$k_Fl$).

 One can see that $\sigma^{DOS}$ and $\sigma^{MT}_{reg}$ give a \emph{negative
contribution} to the fluctuation conductivity, whereas the AL and
the anomalous MT contributions are positive. This leads to a
competition between the positive and the negative contributions.
But as we have shown, at low temperatures, $T \ll T_c$, and strong
magnetic field, $H \gg H_c$, this negative contribution cannot be
compensated by the positive contributions and therefore the entire
fluctuation conductivity is \emph{negative}. The situation holds
even at $T=0$ where the AL and anomalous MT contributions are
equal to zero.
\begin{figure}
    \epsfysize=4cm
        \centerline{\epsfbox{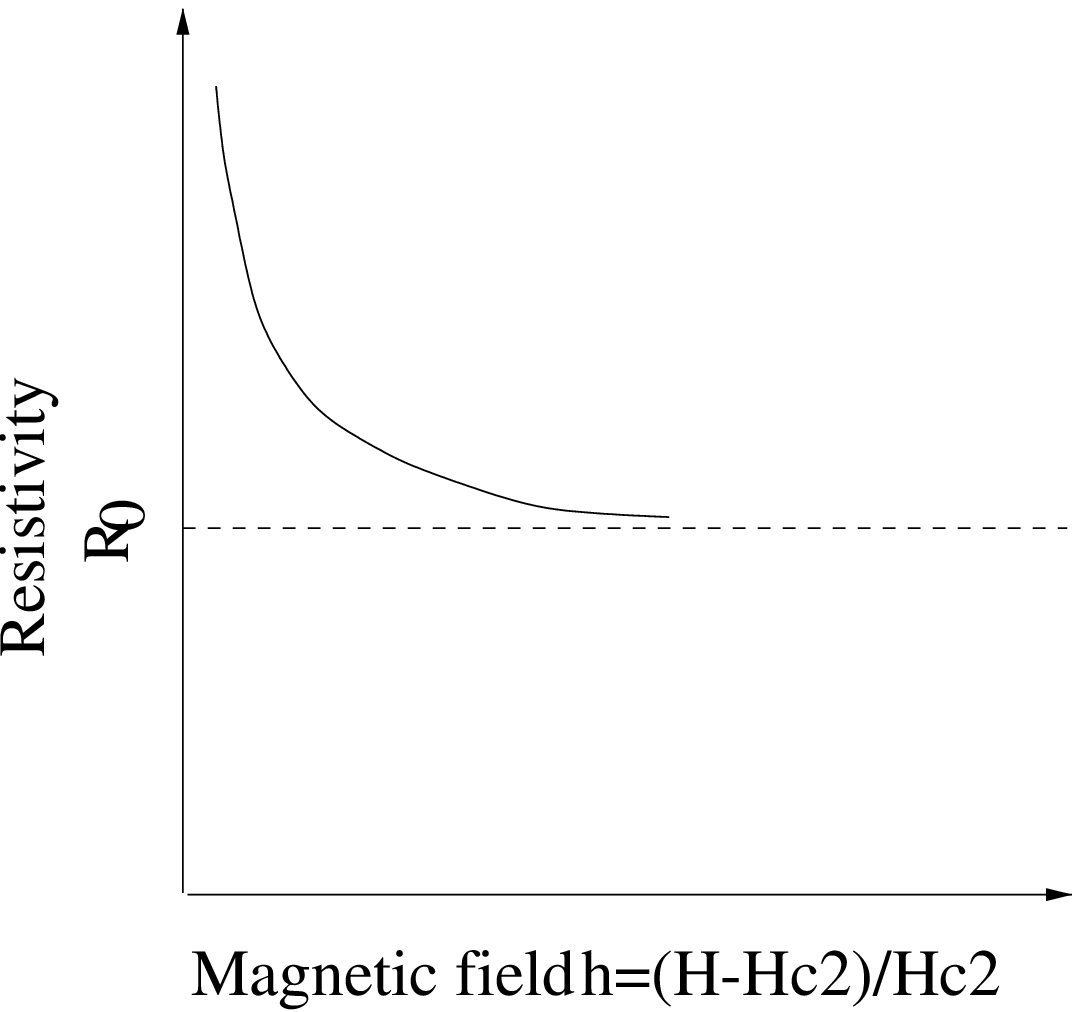}}
        \caption{\label{fig:resistivity} Schematic picture of the resistivity of a
        granulated superconductor at fixed temperature as a function of the magnetic field.}
\end{figure}
Qualitatively results are depicted in Fig.~\ref{fig:resistivity},
where the typical curve for the dependence of the fluctuation
resistivity on the reduced magnetic field $h$ at low temperature
is presented. The curve reaches the value of the classical
resistivity $R_0$ asymptotically only in extremely strong magnetic
fields. It was shown in Ref.~\cite{kn:beloborodov} that for the
granular metals the real transition into the superconducting state
occurs not at $H_c$ but at a lower field $H_{c_2}$, which is due
to the electron motion over many grains. Thus, in order to take
into account the macroscopic orbital electron motion we have to
replace $H_c$ by $H_{c_2}$ in our formulas.

\acknowledgements We acknowledge the support of the SFB 237. The
research of I.~B. was sponsored by the NSF grant DMR-9984002 and
by the A.~P.~Sloan and the Packard Foundations.

\appendix

\renewcommand{\thesection}{\Alph{section}}
\renewcommand{\theequation}{\thesection\arabic{equation}}

\section{} \label{sec:appdos}
\begin{figure}
    \epsfysize=4cm
    \centerline{\epsfbox{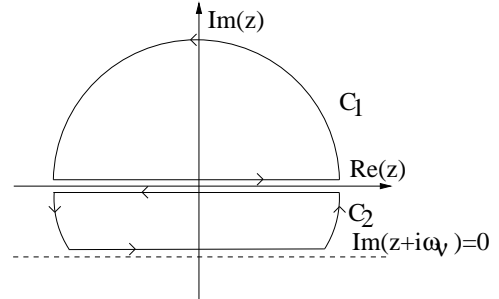}}
    \caption{\label{fig:contourdos} Contours of integration $C_1$ and
    $C_2$, the arrows indicate the way going along the paths.}
\end{figure}

In this appendix we carry out the analytical continuation in
Eq.~(\ref{eq:response}) and obtain the result for the operator of
electromagnetic response $Q^R(\omega)$, Eq.~(\ref{eq:response2}).
The lower limit in the sum over $\Omega_k$ in
Eq.~(\ref{eq:response}) depends on the external frequency
$\omega_{\nu}$. Therefore the analytical continuation of this
expression in the region of real frequencies is not correct by
simply making the replacement $i\omega_{\nu} \rightarrow \omega
+i0^+$, it can be achieved by a transformation of the sum into a
contour integral
\begin{eqnarray} \label{eq:contourintegral}
    \lefteqn{T \sum_{\Omega_k = -\omega_{\nu}}^{\infty}
    D(i\Omega_k,i\omega_{\nu}, {\bf q}) K(i\Omega_k, {\bf q})
    \rightarrow} \nonumber \\
    && \frac{1}{4\pi i}\int_{C_1 + C_2} \coth \frac{z}{2T}
    D(z,i\omega_{\nu}, {\bf q}) K(z, {\bf q}) d z,
\end{eqnarray}
where the contours of integration $C_1$ and $C_2$ are shown in
Fig.~\ref{fig:contourdos}. We should interpret the propagator of
the superconducting fluctuations as the \emph{retarded} propagator
$K^R$ when $\Im (z)>0$ and as the \emph{advanced} propagator $K^A$
when $\Im(z)<0$. In the vicinity of the critical magnetic field,
$(H-H_c)/H_c \ll 1$, we can expand the logarithm in the
denominator of $K$ and the retarded or advanced form of this
quantity has the following form: \cite{kn:beloborodov}
\begin{equation}
    K^{R,A}(-i\Omega, {\bf q})=-\frac{1}{\nu_0}
    \left( 2h \mp \frac{i\Omega}{\Delta_0}+ \eta({\bf q}) \right)^{-1},
\end{equation}
here $h=(H-H_c)/H_c$.
The contribution to the integral from the large circle vanishes if
we extend the contour to $i\infty$, so that only the paths along
the horizontal lines with $\Im(z)=0$ and $\Im(z+i\omega_{\nu})=0$
remain. Thus for the right hand side in
Eq.~(\ref{eq:contourintegral}) we obtain
\begin{eqnarray} \label{eq:analyticalcont}
    &&\int_{-\infty}^{+\infty} \coth
    \frac{z}{2T} D^R(-i z, i\omega_{\nu}, {\bf q})
    K^R(-i z, {\bf q}) \frac{d z}{4\pi i}  \\
    &+& \int_{-\infty}^{\infty} \coth
    \frac{z}{2T} D^A(-i z, i\omega_{\nu}, {\bf q})
    K^A(-i z, {\bf q}) \frac{d z}{4\pi i} \nonumber \\
    &+&  \int_{-\infty-i\omega_{\nu}}^{+\infty-i\omega_{\nu}} \coth
    \frac{z}{2T} D^A(-i z, i\omega_{\nu}, {\bf q})
    K^A(-i z, {\bf q})\frac{d z}{4\pi i}, \nonumber
\end{eqnarray}
where $D^R, D^A$ are the retarded, advanced forms of the function
$D(i\Omega_k, i\omega_\nu, {\bf q})$. They are given by
\begin{eqnarray}
    \lefteqn{D^{R,A}(\Omega, \omega, {\bf q})= -\frac{g e^2}{\pi^2 d T}} \\
    && \times \left[\psi^{\prime} \left( \frac{1}{2} -\frac{2i\omega+i\Omega}{4\pi T} +\alphaq \right)
    -\psi^{\prime} \left(\frac{1}{2} \mp \frac{i\Omega}{4\pi T} + \alphaq \right) \right] \nonumber.
\end{eqnarray}
In the third integral in Eq.~(\ref{eq:analyticalcont}) we make the
substitution of variables $z+i\omega_{\nu} \rightarrow z$ and use
the fact that $\coth \left(\frac{z+i\omega_{\nu}}{2T} \right) =
\coth \left( \frac{z}{2T} \right)$. Now we can simply make the
analytical continuation: $i\omega_\nu \rightarrow \omega$. Finally
we obtain for the sum over $\Omega_k$ in
Eq.~(\ref{eq:contourintegral}) the following result \wide{m}{
\renewcommand{\thesection}{\Alph{section}}
\renewcommand{\theequation}{\thesection\arabic{equation}}
\begin{equation}\label{eq:sumoveromegak2}
     \int_{-\infty}^{+\infty} \frac{d\Omega}{4\pi i} \coth
    \frac{\Omega}{2T} D^R(-i\Omega, \omega, {\bf q}) K^R(-i\Omega,{\bf q})
    +  \int_{-\infty}^{+\infty} \frac{d\Omega}{4\pi i}
    \left( \coth \frac{\Omega-\omega}{2T} - \coth \frac{\Omega}{2T} \right)
    D^A(-i\Omega, \omega, {\bf q}) K^A(-i\Omega, {\bf q}).
\end{equation}
Since we are interested in the dc conductivity it is sufficient to
retain the linear term in $\omega$ in
Eq.~(\ref{eq:sumoveromegak2}) and then we obtain:
\begin{equation} \label{eq:qdos}
    T \sum_{\Omega_k =-\omega_{\nu}}^{\infty} \rightarrow
    i \omega \frac{g e^2}{4 \pi^4 d T^3} \int\limits_{-\infty}^{+\infty}
    \coth \frac{\Omega}{2T} \psi^{\prime \prime}
    \left(\frac{1}{2}-\frac{i\Omega}{4\pi T} +\alphaq \right)
    K^R(-i\Omega, {\bf q}) \frac{d \Omega}{4\pi i}
    -\frac{\omega}{2T} \int\limits_{-\infty}^{+\infty}
    \frac{D^A(-i\Omega,0, {\bf q}) K^A(-i\Omega, {\bf q})}{\sinh^2 \frac{\Omega}{2T}}
    \frac{d \Omega}{4\pi i}.
\end{equation}
} The main contribution to the integral over $\Omega$ in
Eq.~(\ref{eq:qdos}) comes from small values of the frequency
therefore we may put $\Omega=0$ inside $\psi(x)$ in the first
integral and extend the integrand by $i\Omega/ \Delta_0
+\tilde{\eta}$. One can easily see that this integral is
logarithmically divergent and it has to be cut off at
$\Omega_{max} \sim \Delta_0$. In the second integral we make an
expansion in $\Omega$, retaining only the first nonvanishing term
and then we extend the resulting expression by $-i\Omega/ \Delta_0
+\tilde{\eta}$. As a result for the operator of electromagnetic
response we obtain Eq.~(\ref{eq:response2}).

\end{multicols}

\end{document}